# Evidence for Highly p-type doping and type II band alignment in large scale monolayer WSe$_2$ /Se-terminated GaAs heterojunction grown by Molecular beam epitaxy

Debora Pierucci[1], Aymen Mahmoudi[1], Mathieu Silly[2], Federico Bisti[3], Fabrice Oehler[1], Gilles Patriarche[1] Frédéric Bonell[4], Alain Marty[4], Céline Vergnaud[4], Matthieu Jamet[4], Hervé Boukari[5], Emmanuel Lhuillier[6], Marco Pala[1], and Abdelkarim Ouerghi[1]

[1]Université Paris-Saclay, CNRS, Centre de Nanosciences et de Nanotechnologies, 91120, Palaiseau, France
[2]Synchrotron-SOLEIL, Université Paris-Saclay, Saint-Aubin, BP48, F91192 Gif sur Yvette, France
[3]Dipartimento di Scienze Fisiche e Chimiche, Università dell'Aquila, Via Vetoio 10, 67100 L'Aquila, Italy
[4]Université Grenoble Alpes, CNRS, CEA, Grenoble INP, IRIG-Spintec, 38054, Grenoble, France
[5]Université Grenoble Alpes, CNRS and Grenoble INP, Institut Néel, F-38000 Grenoble, France
[6]Sorbonne Université, CNRS, Institut des NanoSciences de Paris, INSP, F-75005 Paris, France.

Two-dimensional materials (2D) arranged in hybrid van der Waals (vdW) heterostructures provide a route toward the assembly of 2D and conventional III-V semiconductors. Here, we report the structural and electronic properties of single layer WSe$_2$ grown by molecular beam epitaxy on Se-terminated GaAs(111)B. Reflection high-energy electron diffraction images exhibit sharp streaky features indicative of a high-quality WSe$_2$ layer produced *via* vdW epitaxy. This is confirmed by in-plane x-ray diffraction. The single layer of WSe$_2$ and the absence of interdiffusion at the interface are confirmed by high resolution X-ray photoemission spectroscopy and high-resolution transmission microscopy. Angle-resolved photoemission investigation revealed a well-defined WSe$_2$ band dispersion and a high p-doping coming from the charge transfer between the WSe$_2$ monolayer and the Se-terminated GaAs substrate. By comparing our results with local and hybrid functionals theoretical calculation, we find that the top of the valence band of the experimental heterostructure is close to the calculations for free standing single layer WSe$_2$. Our experiments demonstrate that the proximity of the Se-terminated GaAs substrate can significantly tune the electronic properties of WSe$_2$. The valence band maximum (VBM, located at the K point of the Brillouin zone) presents an upshifts of about 0.56 eV toward the Fermi level with respect to the VBM of WSe$_2$ on graphene layer, which is indicative of high p-type doping and a key feature for applications in nanoelectronics and optoelectronics.

Two dimensional (2D) materials have attracted much attention thanks to their tunable electronic properties[1–3]. So far, a large body of research has been devoted to evolution of the quasiparticle bandgap as a function of the number of layers, doping and strain in the 2D transition metal dichalcogenides (TMDs)[4-6]. Despite promising physical properties, their integration in optoelectronics devices can be difficult because the fabrication of standard p-n junction requires the control of the carrier density, which is difficult to achieve by substitutional doping with electron acceptors or donors in 2D materials obtained by direct growth[7–10]. The presence of defects, such as chalcogen vacancies, may induce gap states which are responsible for the "as-grown" 2D material doping (i.e., "defect doping"). For example, S vacancies have been reported to act as electron donors in $MoS_2$ and $WS_2$, resulting in a n-type character of the "as-grown" material[11,12]. Similar n-type doping is present in $PtSe_2$ and $WSe_2$ due to the presence of Se vacancies[13,14]. The manipulation of the chalcogen vacancies, *via* hydrogen passivation, provides a route to modulate the carrier concentration of such chalcogen-based monolayers[15,16]. Other attempts to convert n-type "as-grown" TMDs into p-type semiconductors have been reported, *via* the creation of cation vacancies (e.g. Mo or W vacancies)[17] or their substitution with other metal atoms (e.g. Nb atoms or V)[18,19]. Alternatively, p-type doping in TMDs can be achieved using surface adsorption or intercalation of electron-accepting atoms or molecules[20–22]. However, all these techniques present several limitations related to the operational stability of these 2D materials in real devices[23].

Interestingly, the growth of 2D materials on conventional 3D semiconductors results in hybrid 2D/3D heterostructures, which can bring some advantages of the more established 3D semiconductors while retaining some specificities of the 2D materials. An interesting combination is that of GaAs, a mature 3D semiconductor with applications spanning from light emitting diodes to high power electronics[24], with 2D materials such as transition metal mono- and di-chalcogenides[25–27 28]. Although several research works have focused on the growth and fabrication of such heterostructures[29–32], most of the resulting interfacial and electronic properties have not been specifically investigated. Thus, we dedicate this paper to the investigation of the electronic properties of 2D TMD $WSe_2$ and 3D GaAs, with a 2D/3D heterostructure formed at the GaAs(111)B surface[33].

In this work, we report high p-type doping stabilized in the monolayer $WSe_2$ *via* an electron transfer from the $WSe_2$ to the Se-terminated GaAs substrate. A large area high quality crystalline single layer $WSe_2$ film was grown by molecular beam epitaxy (MBE) on Se-terminated GaAs(111)B substrate. The thickness of the $WSe_2$ layer was controlled thanks to a quartz crystal microbalance, by monitoring the amount of deposited W. High-resolution x-ray photoemission spectroscopy (HR-XPS) and high resolution scanning transmission electron microscopy (HR-STEM) investigations indicate that the Se-termination of the GaAs surface weakens the interaction between the TMD film and the substrate[32], resulting in a van der Waals (vdW) epitaxy of $WSe_2$ in absence of any secondary compound detected at the 2D/3D interface. We find that $WSe_2$ grown on Se-terminated GaAs has the same structural properties than monolayer $WSe_2$ obtained through other means. However, we notice a significant upshift of the $WSe_2$ valence band maximum (*i.e. p-type* doping), compared to n-type $WSe_2$ obtained on graphene substrate. Moreover, ARPES measurements reveal that the $WSe_2$/Se-terminated GaAs heterostructure has a type-II band alignment, which is particularly interesting for applications in optoelectronic devices. This experimental study is crucial for understanding such heterostructures in a prospect of real applications, knowing that scaling up devices and processes based on mechanically exfoliated micrometer-sized flakes remains challenging.

**Results and discussions:**

A monolayer thick WSe$_2$ film was grown on a GaAs(111)B substrate by MBE. Prior to the WSe$_2$ growth, the GaAs(111)B surface was passivated with Se in the MBE reactor. During this surface passivation step, Se atoms substitute surface As atoms, bonding covalently with the underlying Ga atoms to form a top uniform Se layer with no dangling bonds[34,35]. The atomic structure of the Se-passivated GaAs(111)B surface and the grown WSe$_2$ layer are presented in figures 1(a)[36,37]. In figure 1 (b), we can show the typical RHEED patterns obtained after the WSe$_2$ growth on the Se-terminated GaAs(111)B surface along two azimuth. The persistence of streaky RHEED patterns after the WSe$_2$ growth indicates a well-ordered and flat epitaxy on GaAs, with a uniform WSe$_2$ layer. The RHEED patterns further indicate that the hexagonal cell of WSe$_2$ aligns with the three-fold surface lattice of GaAs(111)B with [11-20]$_{WSe_2}$//[1-10]$_{GaAs}$ and [1-100]$_{WSe_2}$//[11-2]$_{GaAs}$ in the plane and [0001]$_{WSe_2}$//[-1-1-1]$_{GaAs}$ out of plane. Grazing incidence X-ray diffraction measurements (GIXRD) were performed in order to further characterize the crystallinity of the layers. The incidence angle was set to 0.32°, close to the critical angle of the substrate. In figure 1(c), radial scans along the reciprocal direction of the substrate (h -h 0) labeled R and (2h -h -h) labeled R+30° show diffraction peaks of the substrate and of the WSe$_2$ layer. A radial scan at 15° from the R one, labeled R+15° shows no peaks, giving a first indication that the layers are crystallographically well oriented. From these measurements we obtain the epitaxial relationship [11-20]$_{WSe_2}$//[1-10]$_{GaAs}$. In addition to the well identified WSe$_2$ diffraction peaks, two weak peaks appear in the R+30° scan and one shoulder on the right of the GaAs(2-20) peak in the R scan. The corresponding positions are labeled with "X". The associated diffraction vector lengths are in the ratio 1, 2 and $\sqrt{3}$ respectively, which, considering the 30° angle between the scan directions, corresponds to a hexagonal lattice with a lattice parameter $a_X$=3.86 Å. This could possibly correspond to GaSe layer strongly strained on the GaAs(111) surface. Indeed, bulk GaSe exhibits an hexagonal structure with $a_{GaSe}$=3.75 Å[10] and the GaAs(111) surface plane is hexagonal with a lattice parameter $a_{GaAs-Hex}$=3.998 Å. The value found for $a_X$ lies between the bulk value and the substrate value with a strain of 3% with respect to the bulk. From azimuthal scans at Bragg positions shown in figure S1, the orientation distribution of WSe$_2$ and X layers reach 3.3° and 2.7° respectively, confirming the good epitaxy.

A macroscopic analysis of the chemical composition of the WSe$_2$/Se-terminated GaAs heterostructure was performed using HR-XPS at the synchrotron facility (TEMPO beamline, SOLEIL, France). HR-XPS spectra recorded at a photon energy hν = 120 eV for Se 3p, As 3d, W 4f and Ga 3d are shown in Figure 1(d)-(g), respectively. The spectra were analyzed using the curve fitting procedure described in the Methods section. The experimental data are displayed as black circles. The solid blue line is the envelop of the fitted components. The Ga 3d peak (Figure 1(g)) presents a sharp doublet, characterized by Ga 3d$_{5/2}$ and Ga 3d$_{3/2}$ peaks with a spin orbit splitting (SO) of 0.44 eV and a branching ratio of 1.5. The Ga 3d$_{5/2}$ peak located at a binding energy (BE) of 19.0 eV corresponds to the Ga-As bond in GaAs. The other broad doublet with the Ga 3d$_{5/2}$ peak at BE = 19.6 eV is related to the Ga-Se bond[34] at the interface of the heterostructure. The As 3d peak (Figure 1(e)) is composed of a single doublet (As 3d$_{5/2}$ and As 3d$_{3/2}$ with SO = 0.7 eV and branching ratio of 1.5) at BE$_{As\ 3d5/2}$ = 40.9 eV attributed to As bounded to Ga[38]. In the Se 3d core level (Figure 1(d)), we observe two doublets (Se 3d$_{5/2}$ and Se 3d$_{3/2}$ with SO = 0.86 eV and branching ratio of 1.5). The high intensity doublet Se 3d$_{5/2}$ at BE= 54.1 eV corresponds to the Se atoms embedded in the WSe$_2$ monolayer[39,40], while the second Se 3d$_{5/2}$ at lower BE= 53.4 eV corresponds to the Se bounded to Ga at the interface[41]. The W 4f peak shown in Figure 1(f) presents a main doublet (W 4f $_{7/2}$ and W 4f $_{5/2}$ with a SO= 2.15 eV and branching ratio of 1.3) at BE = 31.8 eV, corresponding to stoichiometric WSe$_2$[40]. We find no additional signature from other compound; *i.e.* no carbon or oxygen related bonds[42,43], which indicates that the surface and the heterostructure are free from contamination. In particular, the absence of covalent As-Se bound or W-Ga bounds points toward a clean and sharp vdW interface between the Se-terminated GaAs(111)B surface and the WSe$_2$ monolayer.

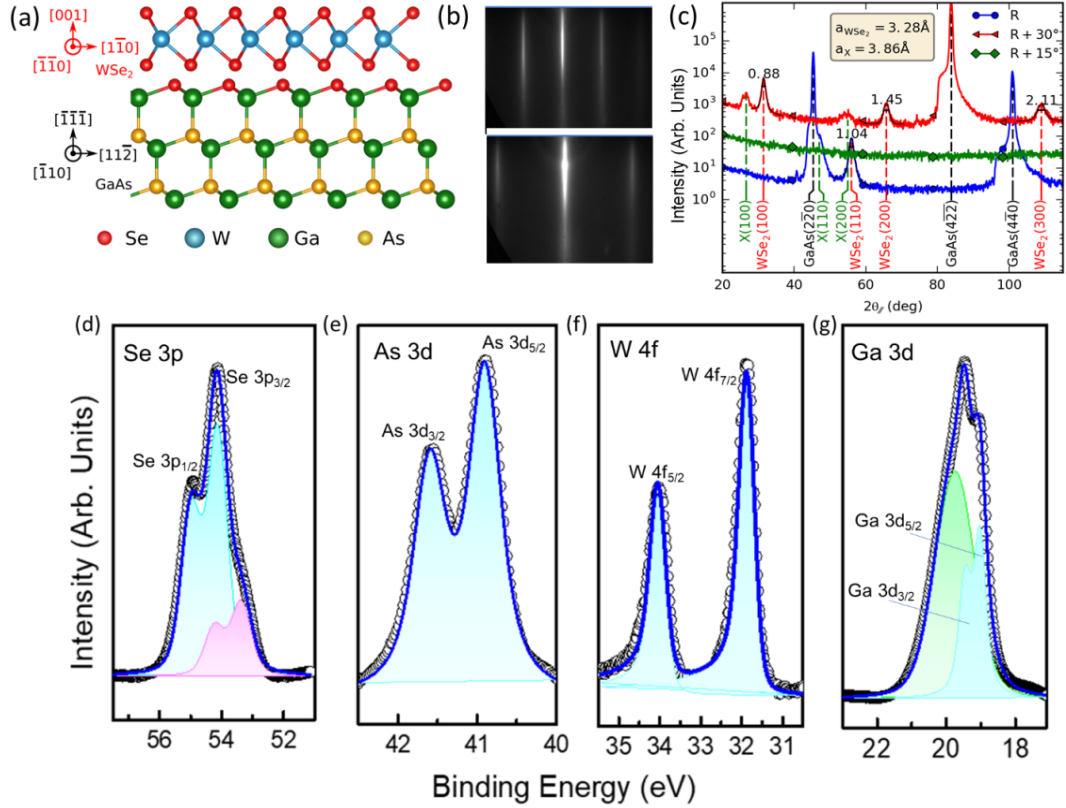

**Figure 1**: Structural and chemical properties of the WSe$_2$/Se-terminated GaAs heterostructure. (a) Schematic structure of a hexagonal WSe$_2$ monolayer on a Se-terminated GaAs(111)B substrate, (b) RHEED patterns of WSe$_2$/ Se-terminated GaAs along two electron beam directions separated by 30°, (c) GIXRD in-plane radial scans at 0°, 30° and 15° from the substrate reference direction R=GaAs(h-h0). The substrate GaAs peaks are indexed in the cubic system whereas the peaks of WSe$_2$ and of an additional "X" layer are indexed in the hexagonal system. The extracted lattice parameters $a_{WSe_2}$ and $a_X$ are given in the insert. The WSe$_2$ peaks are annotated with the full width at half maximum. (d-g) HR-XPS core level spectra of Se-3p, As-3d, W-4f and Ga-3d, respectively.

To confirm the vdW nature of the WSe$_2$/Se-terminated GaAs interface, we have carried out high resolution scanning transmission electron microscopy (HR-STEM). Two cross-sectional views showing atomic resolution are presented in Figure 2(a) and (c), along the [11-2] and [-110] GaAs zone axes, respectively. The corresponding atomic structures of WSe$_2$ and GaAs are overlaid in Figure 2(b) and (d), respectively. The HR-STEM images confirm that the WSe$_2$ film consists of a monolayer with uniform coverage. The atomic resolution allows us to determine the lattice parameter, confirming the expected in-plane [11-20]$_{WSe_2}$//[-110]$_{GaAs}$ and [1-

100]$_{WSe_2}$//[11-2]$_{GaAs}$ and out-of-plane [0001]$_{WSe_2}$ // [-1-1-1]$_{GaAs}$ orientations. The spacing between the bottom Se sub-layer of WSe$_2$ and the top Se-terminated GaAs layer is 3.35 Å, which is very close to the standard vdW gap between consecutive layers in bulk WSe$_2$ (3.37 Å)[44]. Combined with the absence of covalent bond from the XPS data, the spacing from the HR-STEM image points to a vdW bonding at the interface. Therefore, the WSe$_2$/Se-terminated GaAs(111)B heterostructure is indeed a mixed 2D/3D vdW heterostructure. We also observed terraces in the underlying GaAs (111)B surface with clear individual steps (Figure 2(b)). Nevertheless, WSe$_2$ above these steps is continuous and retains a thickness of one monolayer (Figure 2(b)). The atomic structure is resolved on both upper and lower terraces, showing that the WSe$_2$ lattice orientation remains unaffected by the stepped surface. Such arrangement has already been observed in other 2D/3D material heterostructures, such as graphene layers obtained on vicinal SiC(0001) substrates[45]. The persistence of the WSe$_2$ lattice orientation and thickness over a surface which is not perfectly flat indicates the possibility of uniform large scale WSe$_2$ growth using MBE. Such atomically resolved HR-STEM also gives a direct evidence of the "quasi"-vdW gap formed between the Se-terminated GaAs crystal and the 2D layered material.

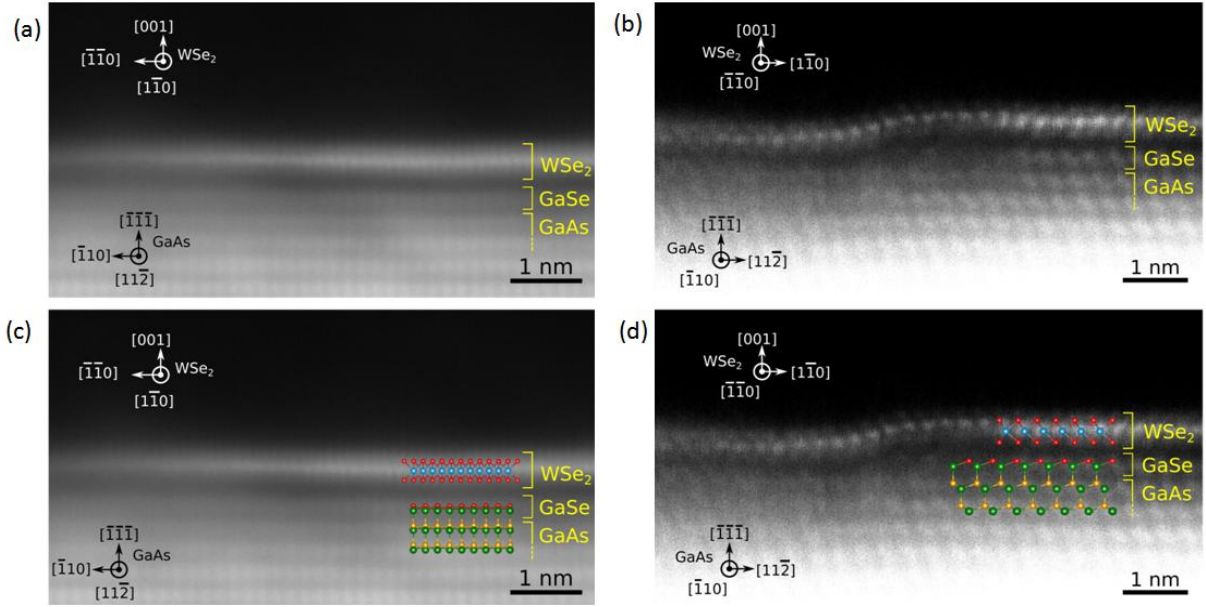

**Figure 2**: Structural investigation of the WSe$_2$/Se-terminated GaAs(111)B heterostructure. (a and b) Atomically resolved bright-field HR-STEM image along two different zone axes, (c and d) Same images with the overlaid atomic structure of the monolayer WSe$_2$ on the Se-terminated GaAs(111)B surface.

Insight into the WSe$_2$ electronic structure was obtained using angle resolved photoemission spectroscopy (ARPES) at the TEMPO beamlines (SOLEIL synchrotron facility, France). In Figure 3(a) and (b), we report the WSe$_2$ and Se-terminated GaAs band structures along the ΓK direction (WSe$_2$) in the Brillouin zone obtained with photon energy of 60 eV and 120 eV, respectively. At low photon energy (60 eV, Figure 3(a)), the ARPES spectrum is dominated by the signature of the topmost WSe$_2$ layer with a valence band maximum (VBM) located at the K point of the Brillouin zone (BE VBM$_{Wse2}$ = -0.57±0.01 eV). At lower binding energy with respect to the VBM, in particular around the Γ point, the signal of a down-dispersing paraboloid related to the GaAs band structure is barely visible. However, at higher photon energy (120 eV, Figure 3(b)), the contribution from the GaAs substrate increases and we now clearly recognize the typical GaAs manifold band of the light-hole (LH), heavy-hole (HH) and spin-orbit split (SO) bands[46–48]. We note that the VBM of GaAs is located at the Γ point, at a BE $VBM_{GaAs}$ = -0.67±0.01 eV (figure S2). In Figure 3(c), we show a superposition of isoenergy cuts of the ΓKM plane obtained in surface sensitive condition (60 eV). The particular isoenergy cut at a BE close to the valence band maximum of WSe$_2$ only reveals electronic states at the K point in all the ΓKM plane. The other isoenergy cuts at lower BE indicate a circular hole pocket at the Γ point and a triangular hole pocket at the K-point. The K pocket exhibits hexagonal symmetry, which indicates that it originates from the hexagonal WSe$_2$ monolayer. The presence of a single pocket at a unique well-determined K position in the plane further confirms that the WSe$_2$ film is single domain and monocrystalline, with a unique lattice alignment between the Se-terminated GaAs(111)B and the WSe$_2$ monolayer in agreement with the RHEED measurements and in-plane x-ray diffraction data.

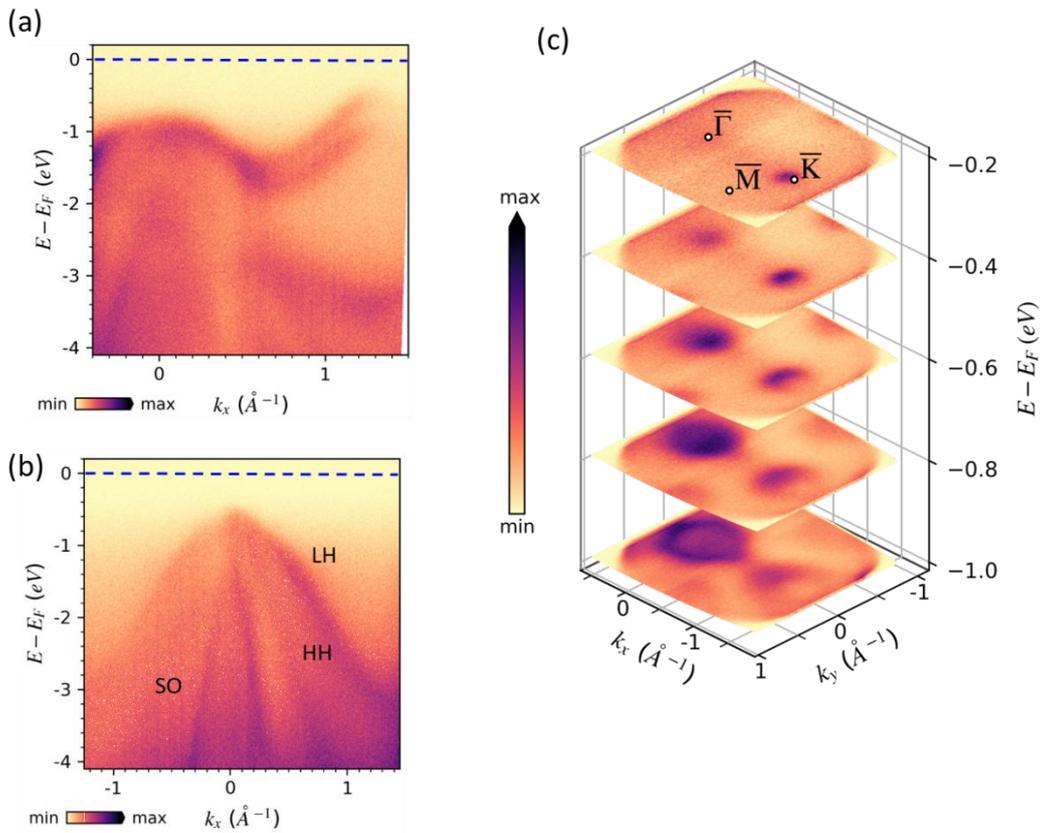

**Figure 3**: ARPES measurements of the WSe$_2$/Se-terminated GaAs heterostructure. (a) Surface-sensitive ARPES measurements of the WSe$_2$/Se-terminated GaAs heterostructure along the ΓK direction of the WSe$_2$ first Brillouin zone measured at a photon energy hν = 60 eV. (b) Corresponding bulk-sensitive ARPES data acquired at hν=120 eV. (c) Iso-energy cuts of the first Brillouin zone of the ΓKM plane obtained in surface-sensitive conditions (hν = 60 eV). The Fermi level position is located at the zero of the binding energy (marked by a blue dotted line).

The clearly visible electronic states and the absence of band hybridization between Se-terminated GaAs and WSe$_2$ indicate a weak interlayer interaction, in line with the vdW nature of the hetero-interface. Although the vdW interaction is weak and does not result in the direct hybridization of the respective electronic bands, it can redistribute the charge density at the interface and create a permanent electric dipole at the WSe$_2$/Se-terminated

GaAs interface. This charge redistribution (or charge transfer) can thus induce a stable doping of the TMD monolayer depending on the substrate. To better clarify this point, the Figure 4 compares the band structure of the above WSe$_2$/Se-terminated GaAs(111)B 2D/3D heterostructure with that of a conventional 2D/2D WSe$_2$/graphene heterostructure. The WSe$_2$/graphene sample consists of a monolayer WSe$_2$ crystal transferred on graphene/SiC substrate (more details of the fabrication process can be found in ref[12]). Unlike the WSe$_2$/Se-terminated GaAs(111)B, for which the WSe$_2$ is grown by MBE, chemical vapor deposition (CVD) is used to obtain the WSe$_2$ crystal of the WSe$_2$/graphene sample. Figure 4 (a) shows the band structure of the WSe$_2$/Se-terminated GaAs(111)B along the ΓK direction, while Figure 4 (b) presents the one of WSe$_2$/graphene in the same orientation. In Figure 4(c) and (d), the corresponding density functional theory (DFT) calculations for a freestanding single layer of WSe$_2$ (blue dotted lines, see the Method section) is overlaid to the ARPES experimental data. The DFT bands are shifted to account for the different Fermi level position. As expected from the weak hybridization in both heterostructures, the top of the valence band near K is essentially composed of states from the WSe$_2$ monolayer. This valence band, is also visible at the top of the Γ point as a single band (here mostly composed of W $d_{z2}$ and Se $p_z$ orbitals) [49], which is another proof that the WSe$_2$ thickness is a monolayer for both heterostructures.

Due to the strong spin-orbit interaction originating from the W $d$ orbitals[50], the band splits into two branches at the K point, where the valence band maximum (VBM) is located. To quantify this splitting, an energy distribution curve (EDC) was extracted at the K point of WSe$_2$. We obtain a spin-orbit coupling (SOC) of 0.45±0.02 eV and 0.47±0.02 eV for WSe$_2$ on Se-terminated GaAs (figure S3) and graphene (figure S4), respectively. This large SOC is intrinsic to WSe$_2$ and mainly originates from W $d_{x^2-y^2} + d_{xy}$ and Se $p_x + p_y$ bonding states. Considering the good agreement between the reference monolayer WSe$_2$ DFT calculations and the experimental ARPES data, these small variations of the SOC further confirm that the WSe$_2$ can be considered as quasi-freestanding on the Se-terminated GaAs(111)B surface, with no interlayer hybridization. Combined with the good agreement with the DFT, we can conclude that the WSe$_2$ band structure in both heterostructures (with VBM$_{WSe2}$ at K) is the same than the one of pristine direct-bandgap monolayer WSe$_2$.

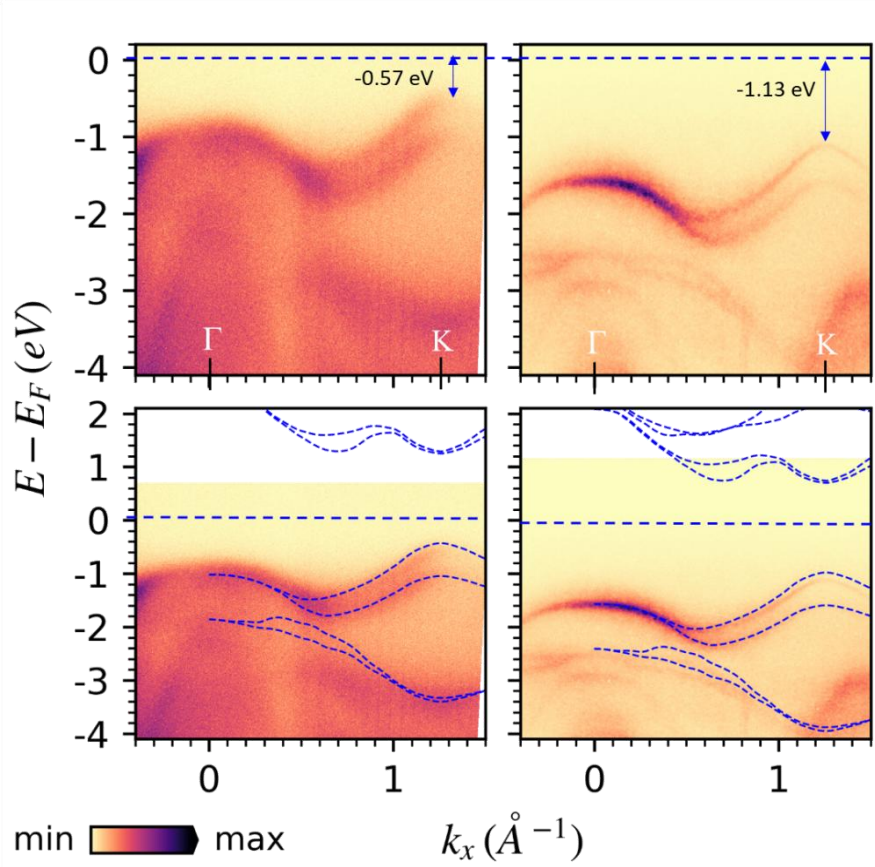

**Figure 4**: Comparison between the experimental band structure of WSe$_2$/Se-terminated GaAs(111)B and WSe$_2$/Graphene heterostructures. (a-b) ARPES measurements of the valence band along the ΓK high symmetry directions of WSe$_2$/Se-terminated GaAs(111)B and single layer WSe$_2$/Graphene, respectively. (c) and (d) Same data with overlaid theoretical DFT calculations from free standing WSe$_2$ monolayer. The Fermi level position is located at the zero of the binding energy (marked by a blue dotted line).

The most striking feature when comparing Figure 4(c) and (d), is large rigid energy shift (*i.e.* simple translation along the y-axis) of the bands with respect to the Fermi level. Such a shift indicates a different charge transfer between the WSe$_2$ and its relative "substrate" between the two heterostructures. This observed charge transfer effectively acts as a substrate-induced doping of the TMD monolayer. In the case of WSe$_2$/graphene the energy position of VBM at the K point is E-E$_F$ = -1.13±0.01 eV. Considering a quasi-particle bandgap of 2.00 eV in monolayer WSe$_2$, the VBM position indicates an electron transfer from the graphene to the WSe$_2$ and a n-type character for majority carriers in the WSe$_2$ monolayer[51]. Conversely, the case of WSe$_2$/ Se-terminated GaAs shows the opposite behavior with a VBM at the K point at energy E-E$_F$ = -0.57±0.01 eV, indicating an electron transfer from the WSe$_2$ to the Se-terminated GaAs and thus a p-type doping of the WSe$_2$. When such charge transfer operates at the interface, its creates a permanent interface dipole, which is characterized by a potential step $\Delta V$[52]. This potential step can be estimated as the difference between the work functions of the pristine Se-terminated GaAs(111)B substrate ($\phi_{GaAs} = 5.2 \pm 0.1\ eV$)[53–57] and of the heterostructure. The latter is experimentally determined *via* the measurement of the secondary electron (SE) edge $\phi_{WSe2/GaAs} = 4.90 \pm 0.05$ eV (Figure S5). We obtain a value of $\Delta V = \phi_{GaAs} - \phi_{WSe2/GaAs} = 0.30 \pm 0.15$ eV, indicating the presence of an electric dipole. On the basis of these results, we are able to determine all the respective band offsets and we can draw the complete electrical schematic of WSe$_2$/Se-terminated GaAs(111)B heterostructure in Figure 5. Considering the known values of the quasiparticle bandgaps for WSe$_2$ (Eg$_{WSe2}$ = 2.00±0.05 eV where the error bar takes into account the variation of the quasiparticle band gap and exciton binding energy induced by different environmental dielectric constants, i.e., different substrates)[58–61] and GaAs (E$_{GaAs}$ = 1.50±0.02 eV, for GaAs we considered the band gap value at low temperature T = 50K)[62,63], we obtain a conduction band discontinuity ΔEc = ΔEv + (Eg$_{WSe2}$- Eg$_{GaAs}$) = 0.60 ± 0.09 eV, where ΔEv = 0.10 ±0.02 eV is the valence band discontinuity. As shown in Figure 5, the final arrangement of the WSe$_2$/Se-terminated GaAs(111)B heterojunction exhibits a staggered gap between WSe$_2$ and GaAs, and qualifies as a type-II band alignment.

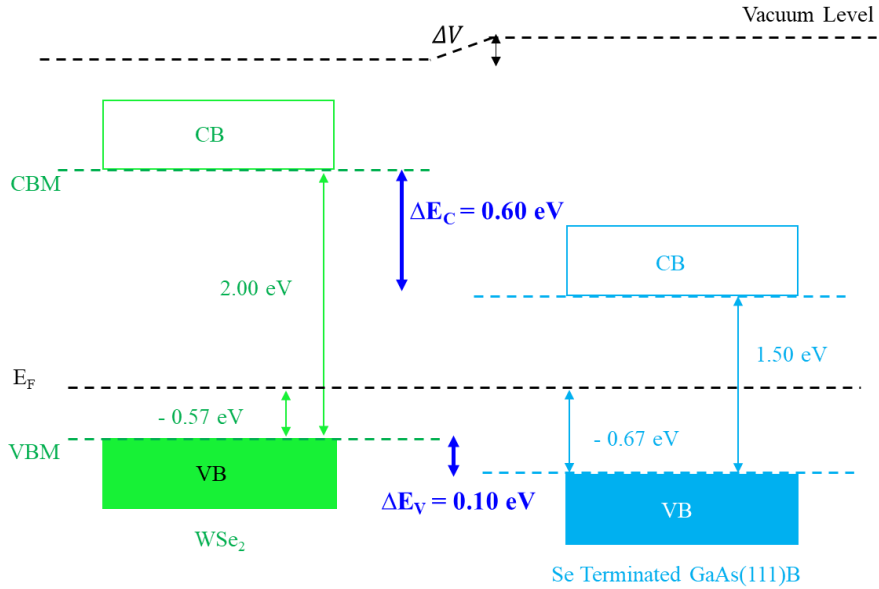

**Figure 5**: Band alignment diagram of the WSe$_2$/Se-terminated GaAs(111)B heterostructure inferred from the HR-XPS and ARPES data. The band alignment is type II, with a staggered forbidden gap between WSe$_2$ and GaAs.

**Conclusion:**

In summary, we have demonstrated the growth of single layer WSe$_2$ on Se-terminated GaAs(111)B using MBE to obtain a mixed vdW 2D/3D heterostructure. The WSe$_2$ monolayer exhibits continuity on terraces and step edges, indicating the possibility of growing large scale TMDs on GaAs using MBE. ARPES measurements revealed that the WSe$_2$ monolayer exhibits a high p-type doping with a VBM located at the K point. We directly observe a strong spin-orbit splitting at the K point of 0.45±0.02 eV, in excellent agreement with the DFT calculations on freestanding material. The ARPES measurements on the heterostructure shows that the WSe$_2$ related bands largely retain their original electronic structure and that freestanding WSe$_2$ monolayer is a good approximation of the electronic properties close to the top of the valence band. In our 2D/3D heterostructure with a sharp van der Waals interface, the majority carriers in the 2D material mostly depends on the respective material band offset. Compared to other fabrication techniques, such a mixed 2D/3D heterostructure offers a natural solution to tailor the majority carrier type and density in the 2D layer. Here, we are able to induce a substantial p-type doping in WSe$_2$, *via* charge transfer to GaAs across the WSe$_2$/Se-terminated GaAs(111)B interface. This first experimental realization of stable and highly doped p-type WSe$_2$ monolayer on a large-area epitaxial GaAs(111)B substrate is likely to apply to other TMDs and 2D materials, to broaden their use in future nanoelectronic and optoelectronic applications in a scalable fashion.

**ACKNOWLEDGMENTS**

We acknowledge the financial support by RhomboG (ANR-17-CE24-0030), MagicValley (ANR-18-CE24-0007) and Graskop (ANR-19-CE09-0026) grants. The LANEF framework (ANR-10-LABX-51-01) is acknowledged for its support with mutualized infrastructure.

**Supporting Information:** In-plane azimuthal XRD scans done with detector positioned at the Bragg angle of $WSe_2(110)$, $WSe_2(100)$ and X(110) Bragg angles as presented in the SI. The band positions were determined by energy curve analysis (EDC) of the ARPES spectra as reported. The work function determination via the measurement of the secondary electron edge is also shown.

**Competing financial interests:** There are no conflicts to declare.

**Author contributions** F.BO., C.V., M.J. and H.B. grew the MBE samples. D.P., M.S. and A.O. carried out the XPS/ARPES measurement. A.M. performed XRD measurements. G.P. and F.O. performed and interpreted TEM. M.P. carried the DFT calculation. D.P., A.O. A.M., E.L., A.M. and F.BI. analyzed the data. All the authors discussed the results and commented on the manuscript.

**Methods:**

**Growth of $WSe_2$/GaAs heterostructure:** A Knudsen cell was used to evaporate Se while an e-beam evaporator was used to evaporate W. The surface of the GaAs(111)B substrate was thermally deoxidized inside the MBE chamber under a $10^{-6}$ mbar Se flux. The same Se flux was used to grow one monolayer of $WSe_2$ at a rate of 0.15 ML/min and at a growth temperature 50°C lower than the one used for the deoxidation of the substrate.

**Grazing incidence X-ray diffraction measurements:** The grazing incidence X-ray diffraction (GIXRD) was done with a SmartLab Rigaku diffractometer equipped with a copper rotating anode beam tube ($K_\alpha$=1.54Å) operating at 45 kV and 200 mA. Parallel in-plane collimators of 0.5° of resolution were used both on the source and detector sides. The diffractometer is equipped with multilayer mirrors on the incident beam and $K_\beta$ filter on the diffracted beam.

**HR-XPS/ARPES measurements:** HR-XPS/ARPES experiments were carried out on the TEMPO beamline (SOLEIL synchrotron facility, France) at low temperature (50K). The photon source was a HU80 Apple II undulator set to deliver linearly polarized light. The photon energy was selected using a high-resolution plane grating monochromator, with a resolving power $E/\Delta E$ that can reach 15,000 on the whole energy range (45 - 1500 eV). During the XPS measurements, the photoelectrons were detected at 0° angle from the sample surface normal $\vec{n}$ and at 46° from the polarization vector $\vec{E}$. The spot size was 100 × 80 (H×V) µm². A Shirley background was subtracted in all core level spectra. The W 4f spectrum was fitted by a sum of a Gaussian function convoluted with a Doniach-Sunjic lineshape. An asymmetry factor α was used, where α = 0.12 eV. The Se 3p, As 3d and Ga 3d spectra were fitted by sums of Voigt curves, *i.e.* the convolution of a Gaussian by a Lorentzian.

**DFT calculations**: Band structures of the monolayer WSe$_2$ were computed within the density functional theory via the QUANTUM EPSRESSO suite. The spin-orbit interaction was considered by addressing non-collinear simulations with fully relativistic pseudopotentials. In order to obtain a better estimation of the electronic bandgap we adopted the HSE hybrid functional to approximate the exchange-correlation term. The cell geometry and atomic positions were fully relaxed with a convergence threshold of $10^{-3}$ (a.u.) for forces and $10^{-4}$ (a.u.) for the energy. Moreover, we used an 10x15x1 Monkhorst-Pack k-point grid and a cutoff energy of 50 Ry, as well as a vacuum space of 22 Å along the vertical direction suppress the interaction with adjacent cells.

**References:**

(1) Novoselov, K. S.; Mishchenko, A.; Carvalho, A.; Castro Neto, A. H. 2D Materials and van Der Waals Heterostructures. *Science (80-. ).* **2016**, *353*, aac9439.

(2) Li, X. Q. and Y. W. and W. L. and J. L. and J.; Qian, X.; Wang, Y.; Li, W.; Lu, J.; Li, J. Modelling of Stacked 2D Materials and Devices. *2D Mater.* **2015**, *2*, 32003.

(3) Bhimanapati, G. R.; Lin, Z.; Meunier, V.; Jung, Y.; Cha, J.; Das, S.; Xiao, D.; Son, Y.; Strano, M. S.; Cooper, V. R.; *et al.* Recent Advances in Two-Dimensional Materials beyond Graphene. *ACS Nano* **2015**, *9*, 11509–11539.

(4) Pierucci, D.; Henck, H.; Naylor, C. H.; Sediri, H.; Lhuillier, E.; Balan, A.; Rault, J. E.; Dappe, Y. J.; Bertran, F.; Le Févre, P.; *et al.* Large Area Molybdenum Disulphide-Epitaxial Graphene Vertical Van Der Waals Heterostructures. *Sci. Rep.* **2016**, *6*, 26656.

(5) Mouri, S.; Miyauchi, Y.; Matsuda, K. Tunable Photoluminescence of Monolayer MoS2 via Chemical Doping. *Nano Lett.* **2013**, *13*, 5944–5948.

(6) Conley, H. J.; Wang, B.; Ziegler, J. I.; Haglund, R. F.; Pantelides, S. T.; Bolotin, K. I. Bandgap Engineering of Strained Monolayer and Bilayer MoS 2. *Nano Lett.* **2013**, *13*, 3626–3630.

(7) Li, X.; Lin, M.-W.; Lin, J.; Huang, B.; Puretzky, A. A.; Ma, C.; Wang, K.; Zhou, W.; Pantelides, S. T.; Chi, M.; *et al.* Two-Dimensional GaSe/MoSe2 Misfit Bilayer Heterojunctions by van Der Waals Epitaxy. *Sci. Adv.* **2016**, *2*, 2:1501882.

(8) Wang, F.; Wang, Z.; Xu, K.; Wang, F.; Wang, Q.; Huang, Y.; Yin, L.; He, J. Tunable GaTe-MoS2 van Der Waals p-n Junctions with Novel Optoelectronic Performance. *Nano Lett.* **2015**, *15*, 7558–7566.

(9) Lee, C.-H.; Lee, G.; van der Zande, A. M.; Chen, W.; Li, Y.; Han, M.; Cui, X.; Arefe, G.; Nuckolls, C.; Heinz, T. F.; *et al.* Atomically Thin P–n Junctions with van Der Waals Heterointerfaces. *Nat. Nanotechnol.* **2014**, *9*, 676–681.

(10) Ben Aziza, Z.; Henck, H.; Pierucci, D.; Silly, M. G.; Lhuillier, E.; Patriarche, G.; Sirotti, F.; Eddrief, M.; Ouerghi, A. Van Der Waals Epitaxy of GaSe/Graphene Heterostructure: Electronic and Interfacial Properties. *ACS Nano* **2016**, *10*, 9679–9686.

(11) Carozo, V.; Wang, Y.; Fujisawa, K.; Carvalho, B. R.; McCreary, A.; Feng, S.; Lin, Z.; Zhou, C.; Perea-López, N.; Elías, A. L.; *et al.* Optical Identification of Sulfur Vacancies: Bound Excitons at the Edges of Monolayer Tungsten Disulfide. *Sci. Adv.* **2017**, *3*.

(12) Pierucci, D.; Henck, H.; Avila, J.; Balan, A.; Naylor, C. H.; Patriarche, G.; Dappe, Y. J.; Silly, M. G.; Sirotti, F.; Johnson, A. T. C.; *et al.* Band Alignment and Minigaps in Monolayer MoS2-Graphene van Der Waals Heterostructures. *Nano Lett.* **2016**, *16*, 4054–4061.

(13) Xu, H.; Zhang, H.; Liu, Y.; Zhang, S.; Sun, Y.; Guo, Z.; Sheng, Y.; Wang, X.; Luo, C.; Wu, X.; *et al.* Controlled Doping of Wafer-Scale PtSe2 Films for Device Application. *Adv. Funct. Mater.* **2019**, *29*.

(14) Tosun, M.; Chan, L.; Amani, M.; Roy, T.; Ahn, G. H.; Taheri, P.; Carraro, C.; Ager, J. W.; Maboudian, R.; Javey, A. Air-Stable n-Doping of WSe2 by Anion Vacancy Formation with Mild Plasma Treatment. *ACS Nano* **2016**, *10*, 6853–6860.

(15) Pierucci, D.; Henck, H.; Aziza, Z. Ben; Naylor, C. H.; Balan, A.; Rault, J. E.; Silly, M. G.; Dappe, Y. J.;

Bertran, F.; Fèvre, P. Le; *et al.* Tunable Doping in Hydrogenated Single Layered Molybdenum Disulfide. *ACS Nano* **2017**, *11*, 1755–1761.

(16) Pallecchi, E.; Lafont, F.; Cavaliere, V.; Schopfer, F.; Mailly, D.; Poirier, W.; Ouerghi, A. High Electron Mobility in Epitaxial Graphene on 4H-SiC(0001) via Post-Growth Annealing under Hydrogen. *Sci. Rep.* **2014**, *4*.

(17) Seo, S.-Y.; Park, J.; Park, J.; Song, K.; Cha, S.; Sim, S.; Choi, S.-Y.; Yeom, H. W.; Choi, H.; Jo, M.-H. Writing Monolithic Integrated Circuits on a Two-Dimensional Semiconductor with a Scanning Light Probe. *Nat. Electron. 2018 19* **2018**, *1*, 512–517.

(18) Suh, J.; Park, T.-E.; Lin, D.-Y.; Fu, D.; Park, J.; Hee, ‖; Jung, J.; Chen, Y.; Ko, C.; Jang, C.; *et al.* Doping against the Native Propensity of MoS 2 : Degenerate Hole Doping by Cation Substitution. **2014**.

(19) Mallet, P.; Chiapello, F.; Okuno, H.; Boukari, H.; Jamet, M.; Veuillen, J. Y. Bound Hole States Associated to Individual Vanadium Atoms Incorporated into Monolayer WSe2. *Phys. Rev. Lett.* **2020**, *125*, 1–6.

(20) Kang, D. H.; Kim, M. S.; Shim, J.; Jeon, J.; Park, H. Y.; Jung, W. S.; Yu, H. Y.; Pang, C. H.; Lee, S.; Park, J. H. High-Performance Transition Metal Dichalcogenide Photodetectors Enhanced by Self-Assembled Monolayer Doping. *Adv. Funct. Mater.* **2015**, *25*, 4219–4227.

(21) Li, Y.; Xu, C.-Y.; Hu, P.; Zhen, L. Carrier Control of MoS2 Nanoflakes by Functional Self-Assembled Monolayers. *ACS Nano* **2013**, *7*, 7795–7804.

(22) Chen, C.-H.; Wu, C.-L.; Pu, J.; Chiu, M.-H.; Kumar, P.; Takenobu, T.; Li, L.-J. Hole Mobility Enhancement and p -Doping in Monolayer WSe2 by Gold Decoration. *2D Mater.* **2014**, *1*, 034001.

(23) Yoo, H.; Heo, K.; Ansari, M. H. R.; Cho, S. Recent Advances in Electrical Doping of 2D Semiconductor Materials: Methods, Analyses, and Applications. *Nanomater. 2021, Vol. 11, Page 832* **2021**, *11*, 832.

(24) Kang, M. S.; Lee, C. H.; Park, J. B.; Yoo, H.; Yi, G. C. Gallium Nitride Nanostructures for Light-Emitting Diode Applications. *Nano Energy* **2012**, *1*, 391–400.

(25) Ruzmetov, D.; Zhang, K.; Stan, G.; Kalanyan, B.; Bhimanapati, G. R.; Eichfeld, S. M.; Burke, R. A.; Shah, P. B.; O'Regan, T. P.; Crowne, F. J.; *et al.* Vertical 2D/3D Semiconductor Heterostructures Based on Epitaxial Molybdenum Disulfide and Gallium Nitride. *ACS Nano* **2016**, *10*, 3580–3588.

(26) Lee, C. H.; Krishnamoorthy, S.; O'Hara, D. J.; Johnson, J. M.; Jamison, J.; Myers, R. C.; Kawakami, R. K.; Hwang, J.; Rajan, S. Molecular Beam Epitaxy of 2D-Layered Gallium Selenide on GaN Substrates. *arXiv:1610.06265* **2016**.

(27) Gan, L. M.; Diodes, P. N.; Ii, E. W. L.; Lee, C. H.; Paul, P. K.; Ma, L.; Mcculloch, W. D. Layer-Transferred MoS2/GaN PN Diodes. *arXiv:1505.05196* **2015**.

(28) Jariwala, D.; Marks, T. J.; Hersam, M. C. Mixed-Dimensional van Der Waals Heterostructures. *Nat. Mater.* **2016**, *16*, 170–181.

(29) Vishwanath, S.; Sundar, A.; Liu, X.; Azcatl, A.; Lochocki, E.; Woll, A. R.; Rouvimov, S.; Hwang, W. S.; Lu, N.; Peng, X.; *et al.* MBE Growth of Few-Layer 2H-MoTe2 on 3D Substrates. *J. Cryst. Growth* **2018**, *482*, 61–69.

(30) Chen, M.-W.; Ovchinnikov, D.; Lazar, S.; Pizzochero, M.; Whitwick, M. B.; Surrente, A.; Baranowski, M.; Sanchez, O. L.; Gillet, P.; Plochocka, P.; *et al.* Highly Oriented Atomically Thin Ambipolar MoSe2 Grown by Molecular Beam Epitaxy. *ACS Nano* **2017**, *11*, 6355–6361.

(31) Ohtake, A.; Sakuma, Y. Two-Dimensional WSe2/MoSe2 Heterostructures Grown by Molecular-Beam Epitaxy. *J. Phys. Chem. C* **2021**, *125*, 11257–11261.

(32) Ohtake, A.; Sakuma, Y. Evolution of Surface and Interface Structures in Molecular-Beam Epitaxy of MoSe2 on GaAs(111)A and (111)B. *Cryst. Growth Des.* **2016**, *17*, 363–367.

(33) Ogorzalek, Z.; Seredynski, B.; Kret, S.; Kwiatkowski, A.; Korona, K. P.; Grzeszczyk, M.; Mierzejewski, J.; Wasik, D.; Pacuski, W.; Sadowski, J.; *et al.* Charge Transport in MBE-Grown 2H-MoTe2bilayers with Enhanced Stability Provided by an AlO: Xcapping Layer. *Nanoscale* **2020**, *12*, 16535–16542.

(34) Scimeca, T.; Watanabe, Y.; Maeda, F.; Berrigan, R.; Oshima, M. Controlled Passivation of GaAs by Se Treatment. *Appl. Phys. Lett.* **1993**, *62*, 1667–1669.

(35) Ohtake, A.; Goto, S.; Nakamura, J. Atomic Structure and Passivated Nature of the Se-Treated GaAs(111)B Surface. *Sci. Rep.* **2018**, *8*, 1–8.

(36) Gong, Y.; Lin, J.; Wang, X.; Shi, G.; Lei, S.; Lin, Z.; Zou, X.; Ye, G.; Vajtai, R.; Yakobson, B. I.; *et al.* Vertical and In-Plane Heterostructures from WS2/MoS2 Monolayers. *Nat. Mater.* **2014**, *13*, 1135–1142.

(37) Li, B.; Huang, L.; Zhong, M.; Li, Y.; Wang, Y.; Li, J.; Wei, Z. Direct Vapor Phase Growth and Optoelectronic Application of Large Band Offset SnS2/MoS2 Vertical Bilayer Heterostructures with High Lattice Mismatch. *Adv. Electron. Mater.* **2016**, *2*, 1600298.

(38) Ouerghi, a.; Marangolo, M.; Eddrief, M.; Lipinski, B.; Etgens, V.; Lazzeri, M.; Cruguel, H.; Sirotti, F.; Coati, a.; Garreau, Y. Surface Reconstructions of Epitaxial MnAs Films Grown on GaAs(111)B. *Phys. Rev. B* **2006**, *74*, 155412.

(39) Boscher, N. D.; Carmalt, C. J.; Parkin, I. P. Atmospheric Pressure Chemical Vapor Deposition of WSe2 Thin Films on Glass - Highly Hydrophobic Sticky Surfaces. *J. Mater. Chem.* **2006**, *16*, 122–127.

(40) Lin, Y.-C.; Ghosh, R. K.; Addou, R.; Lu, N.; Eichfeld, S. M.; Zhu, H.; Li, M.-Y.; Peng, X.; Kim, M. J.; Li, L.-J.; *et al.* Atomically Thin Resonant Tunnel Diodes Built from Synthetic van Der Waals Heterostructures. *Nat. Commun.* **2015**, *6*, 7311.

(41) González, C.; Benito, I.; Ortega, J.; Jurczyszyn, L.; Blanco, J. M.; Pérez, R.; Flores, F.; Kampen, T. U.; Zahn, D. R. T.; Braun, W. Selenium Passivation of GaAs(001): A Combined Experimental and Theoretical Study. *J. Phys. Condens. Matter* **2004**, *16*, 2187–2206.

(42) Osherov, A.; Matmor, M.; Froumin, N.; Ashkenasy, N.; Golan, Y. Surface Termination Control in Chemically Deposited PbS Films: Nucleation and Growth on GaAs(111)A and GaAs(111)B. *J. Phys. Chem. C* **2011**, *115*, 16501–16508.

(43) Surdu-Bob, C. C.; Saied, S. O.; Sullivan, J. L. An X-Ray Photoelectron Spectroscopy Study of the Oxides of GaAs. *Appl. Surf. Sci.* **2001**, *183*, 126–136.

(44) Schutte, W. J.; De Boer, J. L.; Jellinek, F. Crystal Structures of Tungsten Disulfide and Diselenide. *J. Solid State Chem.* **1987**, *70*, 207–209.

(45) Ouerghi, A.; Silly, M. G.; Marangolo, M.; Mathieu, C.; Eddrief, M.; Picher, M.; Sirotti, F.; El Moussaoui, S.; Belkhou, R. Large-Area and High-Quality Epitaxial Graphene on off-Axis Sic Wafers. *ACS Nano* **2012**, *6*.

(46) Kobayashi, M.; Muneta, I.; Schmitt, T.; Patthey, L.; Ohya, S.; Tanaka, M.; Oshima, M.; Strocov, V. N. Digging up Bulk Band Dispersion Buried under a Passivation Layer. *Appl. Phys. Lett.* **2012**, *101*, 242103.

(47) Cardona, M.; Christensen, N. E.; Fasol, G. Relativistic Band Structure and Spin-Orbit Splitting of Zinc-Blende-Type Semiconductors. *Phys. Rev. B* **1988**, *38*, 1806.

(48) Luo, J. W.; Bester, G.; Zunger, A. Full-Zone Spin Splitting for Electrons and Holes in Bulk GaAs and GaSb. *Phys. Rev. Lett.* **2009**, *102*, 056405.

(49) Dendzik, M.; Michiardi, M.; Sanders, C.; Bianchi, M.; Miwa, J. A.; Gr??nborg, S. S.; Lauritsen, J. V.; Bruix, A.; Hammer, B.; Hofmann, P. Growth and Electronic Structure of Epitaxial Single-Layer WS2 on Au(111). *Phys. Rev. B - Condens. Matter Mater. Phys.* **2015**, *92*, 1–7.

(50) Fan, X.; Singh, D. J.; Zheng, W. Valence Band Splitting on Multilayer $MoS_2$ : Mixing of Spin–Orbit Coupling and Interlayer Coupling. *J. Phys. Chem. Lett.* **2016**, *7*, 2175–2181.

(51) Dappe, Y. J.; Almadori, Y.; Dau, M. T.; Vergnaud, C.; Jamet, M.; Paillet, C.; Journot, T.; Hyot, B.; Pochet, P.; Grévin, B. Charge Transfers and Charged Defects in WSe2/Graphene-SiC Interfaces. *Nanotechnology* **2020**, *31*, 255709.

(52) Si, C.; Lin, Z.; Zhou, J.; Sun, Z. Controllable Schottky Barrier in GaSe/Graphene Heterostructure: The Role of Interface Dipole. *2D Mater.* **2016**, *4*, 015027.

(53) Vitomirov, I. M.; Raisanen, A.; Finnefrock, A. C.; Viturro, R. E.; Brillson, L. J.; Kirchner, P. D.; Pettit, G. D.; Woodall, J. M. Geometric Ordering, Surface Chemistry, Band Bending, and Work Function at Decapped GaAs(100) Surfaces. *Phys. Rev. B* **1992**, *46*, 13293.

(54) Suzuki, S.; Maeda, F.; Watanabe, Y.; Ohno, T. Work Function Changes of GaAs Surfaces Induced by Se Treatment. *Japanese J. Appl. Physics, Part 1 Regul. Pap. Short Notes Rev. Pap.* **1999**, *38*, 5847–5850.

(55) Schindler, P.; Riley, D. C.; Bargatin, I.; Sahasrabuddhe, K.; Schwede, J. W.; Sun, S.; Pianetta, P.; Shen, Z. X.; Howe, R. T.; Melosh, N. A. Surface Photovoltage-Induced Ultralow Work Function Material for Thermionic Energy Converters. *ACS Energy Lett.* **2019**, 2436–2443.

(56) Rumaner, L. E.; Olmstead, M. A.; Ohuchi, F. S. Interaction of GaSe with GaAs(111): Formation of Heterostructures with Large Lattice Mismatch. *J. Vac. Sci. Technol. B Microelectron. Nanom. Struct. Process. Meas. Phenom.* **1998**, *16*, 977.

(57) Heun, S.; Watanabe, Y.; Ressel, B.; Schmidt, T.; Prince, K. C. Valence Band Alignment and Work Function of Heteroepitaxial Nanocrystals on GaAs(001). *J. Vac. Sci. Technol. B Microelectron. Nanom. Struct. Process. Meas. Phenom.* **2001**, *19*, 2057.

(58) He, K.; Kumar, N.; Zhao, L.; Wang, Z.; Mak, K. F.; Zhao, H.; Shan, J. Tightly Bound Excitons in Monolayer WSe2. *Phys. Rev. Lett.* **2014**, *113*, 026803.

(59) Chiu, M.-H.; Zhang, C.; Shiu, H.-W.; Chuu, C.-P.; Chen, C.-H.; Chang, C.-Y. S.; Chen, C.-H.; Chou, M.-Y.; Shih, C.-K.; Li, L.-J. Determination of Band Alignment in the Single-Layer MoS2/WSe2 Heterojunction. *Nat. Commun.* **2015**, *6*, 7666.

(60) Zribi, J.; Khalil, L.; Zheng, B.; Avila, J.; Pierucci, D.; Brulé, T.; Chaste, J.; Lhuillier, E.; Asensio, M. C.; Pan, A.; *et al.* Strong Interlayer Hybridization in the Aligned SnS2/WSe2 Hetero-Bilayer Structure. *npj 2D Mater. Appl.* **2019**, *3*, 27.

(61) Hsu, W. T.; Quan, J.; Wang, C. Y.; Lu, L. S.; Campbell, M.; Chang, W. H.; Li, L. J.; Li, X.; Shih, C. K. Dielectric Impact on Exciton Binding Energy and Quasiparticle Bandgap in Monolayer WS2 and WSe2. *2D Mater.* **2019**, *6*, 025028.

(62) Ehrenreich, H. Band Structure and Electron Transport of GaAs. *Phys. Rev.* **1960**, *120*, 1951–1963.

(63) Nascimento, L. N.; Jamshidi, L. C. L. A.; Barbosa, C. M. B. de M.; Rodbari, R. J. SEMICONDUCTORS OF CRYSTALLIN ALLOYS IN SUPERLATTICES. *Rev. Tecnológica* **2015**, *24*, 81–93.

# Evidence for Highly p-type doping and type II band alignment in large scale monolayer WSe$_2$ /Se-terminated GaAs heterojunction grown by Molecular beam epitaxy

Debora Pierucci[1], Aymen Mahmoudi[1], Mathieu Silly[2], Federico Bisti[3], Fabrice Oehler[1], Gilles Patriarche[1] Frédéric Bonell[4], Alain Marty[4], Céline Vergnaud[4], Matthieu Jamet[4], Hervé Boukari[5], Marco Pala[1], and Abdelkarim Ouerghi[1]

[1] Université Paris-Saclay, CNRS, Centre de Nanosciences et de Nanotechnologies, 91120, Palaiseau, [2]Synchrotron-SOLEIL, Université Paris-Saclay, Saint-Aubin, BP48, F91192 Gif sur Yvette, France
[3]Dipartimento di Scienze Fisiche e Chimiche, Università dell'Aquila, Via Vetoio 10, 67100 L'Aquila, Italy
[4]Université Grenoble Alpes, CEA, CNRS, Grenoble INP, IRIG-Spintec, 38054, Grenoble, France
[5] Université Grenoble Alpes, CNRS and Grenoble INP, Institut Néel, F-38000 Grenoble, France

**Azimuthal XRD scans at Bragg positions:**

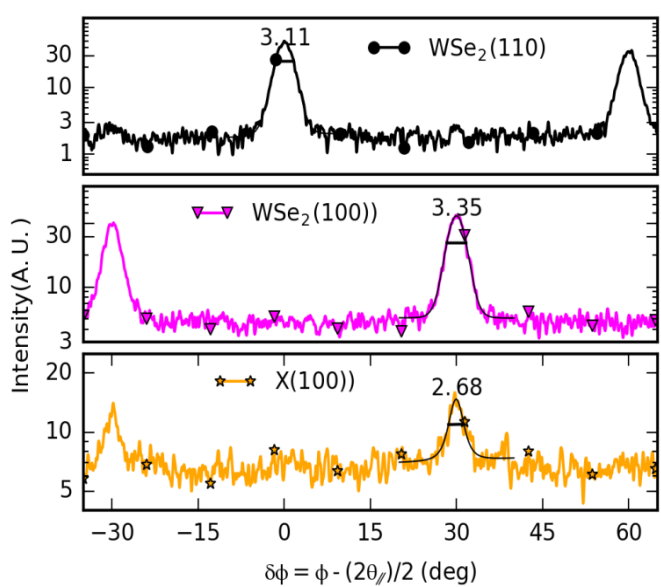

**Figure S1**: In-plane azimuthal XRD scans done with detector positioned at the Bragg angle of WSe$_2$(110), WSe$_2$(100) and X(110) Bragg angles. The peaks are annotated with their full width at half maximum which give the orientation distribution spread of the crystallites.

**Energy Distribution Curves (EDC) analysis:** The band positions were determined by energy distribution curve analysis (EDC) of the ARPES spectra as reported in Figures S2-S4. The fit was performed using a Shirley background (arctangent function as background) and a Gaussian (Lorentzian) line shape for the peaks in the case of K points of WSe$_2$ on Se-terminated GaAs (Graphene), and for the Γ point of Se-terminated GaAs.

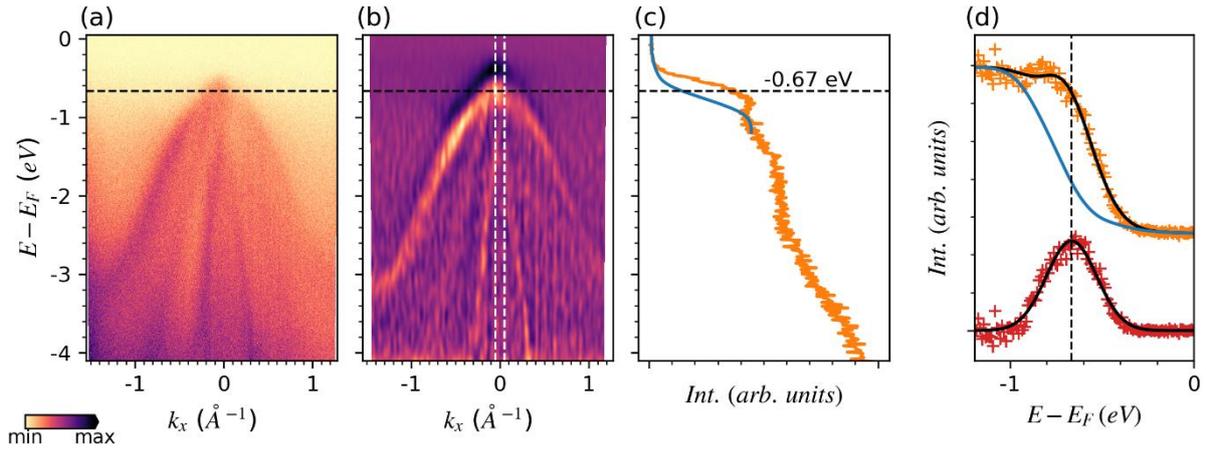

**Figure S2:** Energy position determination for the lowest bonded GaAs band at Γ. (a) the ARPES spectra of WSe$_2$/Se-terminated GaAs probed with 120 eV photon energy; (b) the filtered ARPES spectra using a Laplacian of Gaussian filter; white dashed line indicates the region of integration for the EDC; (c) the EDC signal (in orange) with the Shirley background (blue line); (d) the EDC signal (in orange), the Shirley background (blue line), the background subtracted EDC signal (in red), the fitted line shape (black lines). Dashed black lines in all the panels indicate the energy of the peak position.

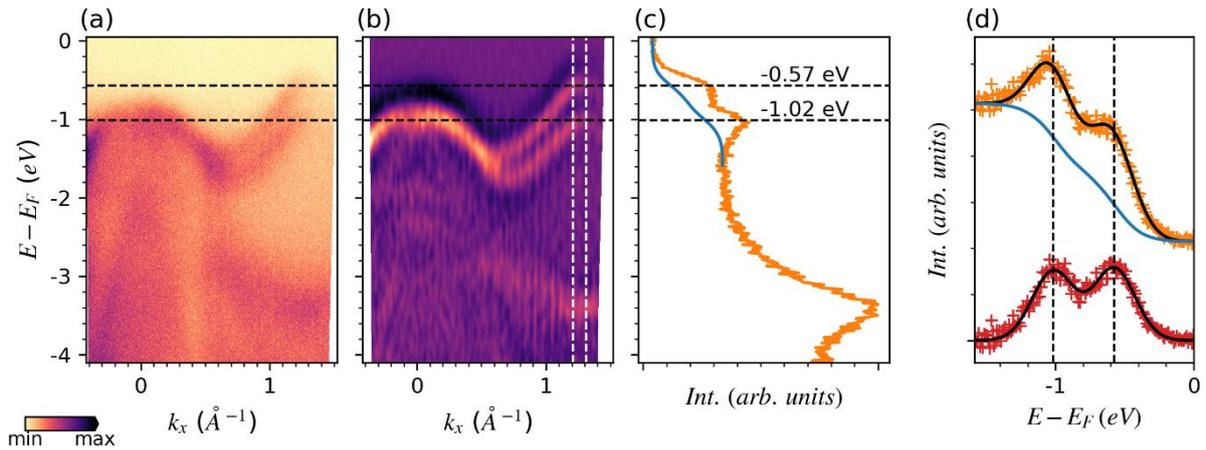

**Figure S3**: Energy position determination for the two lowest bonded WSe$_2$ band at K. (a) the ARPES spectra of WSe$_2$/ Se-terminated GaAs probed with 60 eV photon energy; (b) the filtered ARPES spectra using a Laplacian of Gaussian filter; white dashed line indicates the region of integration for the EDC; (c) the EDC signal (in orange) with the Shirley background (blue line); (d) the EDC signal (in orange), the Shirley background (blue line), the background subtracted EDC signal (in red), the fitted line shape (black lines). Dashed black lines in all the panels indicate the energy of the peak position.

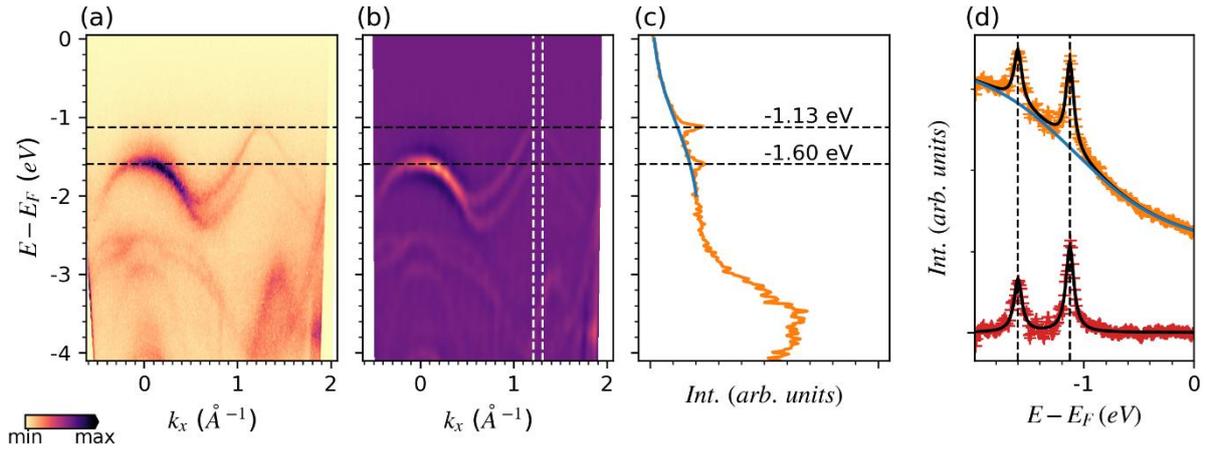

**Figure S4:** Energy position determination for the two lowest bonded WSe$_2$ band at K. (a) the ARPES spectra of WSe$_2$/graphene; (b) the filtered ARPES spectra using a Laplacian of Gaussian filter; white dashed line indicates the region of integration for the EDC; (c) the EDC signal (in orange) with the arctangent function as background (blue line); (d) the EDC signal (in orange), the arctangent function as background (blue line), the background subtracted EDC signal (in red), the fitted line shape (black lines). Dashed black lines in all the panels indicate the energy of the peak position.

**Work Function measurements:** The work function (WF) of the sample was determined by the measurement of the secondary electron (SE) edge with a photon energy of 120 eV. In order to ensure that the SE has a kinetic energy (KE) higher than the analyser vacuum level, the sample is negatively biased (-18 V) with respect to the analyser. The secondary electron energy distribution curve is shown in Figure S5 as a function of the kinetic energy referenced at the Fermi level. Consequenly, the secondary electron cut-off (KE$_{cutoff}$), measured by extrapolating the edge of the peak to the zero baseline (Figure S5) gives directly the value of the workfunction, which is found to be 4.90 ± 0.05 eV.

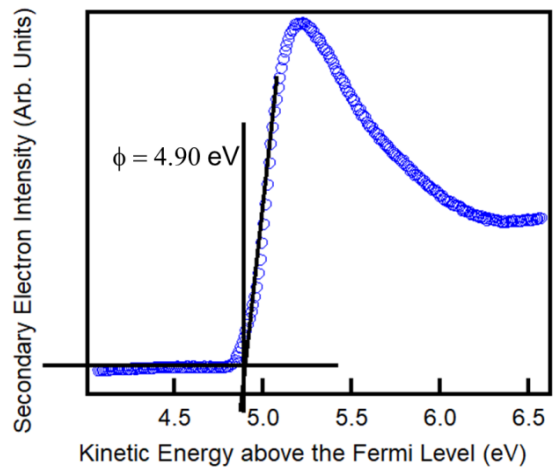

**Figure S5:** Secondary electron intensity as a function of the kinetic energy above the Fermi level for the WSe$_2$/Se-terminated GaAs heterostructure measured with hν = 120 eV.